 \documentclass[pmlr,twocolumn,10pt]{jmlr}

\usepackage{booktabs}
\usepackage{siunitx}

\theorembodyfont{\upshape}
\theoremheaderfont{\scshape}
\theorempostheader{:}
\theoremsep{\newline}

\jmlrvolume{225}
\jmlryear{2023}
\jmlrsubmitted{LEAVE UNSET}
\jmlrpublished{LEAVE UNSET}
\jmlrworkshop{Machine Learning for Health (ML4H) 2023}

 \title[Learning Generalized Medical Image Representations through Image-Graph Contrastive Pretraining]{Learning Generalized Medical Image Representations Through Image-Graph Contrastive Pretraining}

\author{
    \Name{Sameer Khanna} \Email{sameer\_khanna@berkeley.edu}\\
    \Name{Daniel Michael} \Email{danieljm@cs.stanford.edu }\\
    \addr Department of Computer Science, Stanford University \\ \\
    \Name{Marinka Zitnik} \Email{marinka@hms.harvard.edu}\\
    \Name{Pranav Rajpurkar} \Email{pranav\_rajpurkar@hms.harvard.edu }\\
    \addr Department of Biomedical Informatics, Harvard Medical School\\
}

\begin{document}

\maketitle

\begin{abstract}
Medical image interpretation using deep learning has shown promise but often requires extensive expert-annotated datasets. To reduce this annotation burden, we develop an Image-Graph Contrastive Learning framework that pairs chest X-rays with structured report knowledge graphs automatically extracted from radiology notes. Our approach uniquely encodes the disconnected graph components via a relational graph convolution network and transformer attention. In experiments on the CheXpert dataset, this novel graph encoding strategy enabled the framework to outperform existing methods that use image-text contrastive learning in 1\% linear evaluation and few-shot settings, while achieving comparable performance to radiologists. By exploiting unlabeled paired images and text, our framework demonstrates the potential of structured clinical insights to enhance contrastive learning for medical images. This work points toward reducing demands on medical experts for annotations, improving diagnostic precision, and advancing patient care through robust medical image understanding.
\end{abstract}

\section{Introduction} 

Medical image interpretation is essential for diagnosing and guiding treatment plans across a variety of health conditions, including tasks like chest x-ray analysis. Recent advances in deep learning have exhibited remarkable capabilities in this domain, with some models even matching or exceeding the performance of medical experts \citep{10.1167/iovs.16-19964, shih2019augmenting, wang2020covid}. However, these advances are tempered by a significant bottleneck: the requirement for large, high-quality, labeled datasets for training. The manual annotation process is laborious and resource-consuming, often requiring extensive clinician involvement to label hundreds of thousands of images \citep{de2018clinically, rajpurkar2018deep, rajpurkar2020chexaid}.

The ideal scenario would involve training robust deep learning systems with far fewer labeled examples, thus reducing the annotation burden from hundreds of thousands to just thousands of images. Innovative methods such as transfer learning and self-supervised techniques like contrastive learning have shown promise in enabling more efficient training with less labeled data \citep{irvin2019chexpert, ke2021chextransfer, chen2019self}. Contrastive learning, which distinguishes between similar and dissimilar data pairs, has been successfully applied in medical imaging to augment both image-only and image-text approaches \citep{chen2020simple, misra2020self, radford2021learning}.

While image-text contrastive learning that pairs medical images with free-text radiology reports has demonstrated particular promise in disease detection, its performance remains sub-optimal due to the complex and variable nature of radiology reports \citep{endo2021retrieval}.

To address these challenges, we introduce a novel approach: Image-Graph Contrastive Learning (IGCL). IGCL leverages structured knowledge graphs representing radiology reports, as opposed to free-text reports \citep{radgraph-jain}. This enables the model to focus on key entities and relations, effectively filtering out stylistic variations and thus facilitating more uniform training. To the best of our knowledge, this is the first time an image-graph pairing technique has been proposed in this context.

Standard graph encoders proved inadequate for handling medical report graphs with multiple disconnected components. To effectively encode these fragmented but critical pieces of information, we propose a unique architecture combining a Relational Graph Convolutional Network (RGCN) \citep{schlichtkrull2018modeling} and a transformer encoder \citep{vaswani2017attention}.

Our rigorous evaluations on tasks using the CheXpert dataset reveal that IGCL significantly outperforms existing contrastive pre-training approaches, even when very limited labeled data are available. These findings underscore the potential of IGCL to substantially mitigate the need for manual annotation, thereby advancing medical AI through structured knowledge graphs with broad implications for real-world medical diagnosis and treatment planning.

\begin{figure*}
  \label{fig:overview}
  \centering
  \includegraphics[width=0.8\linewidth]{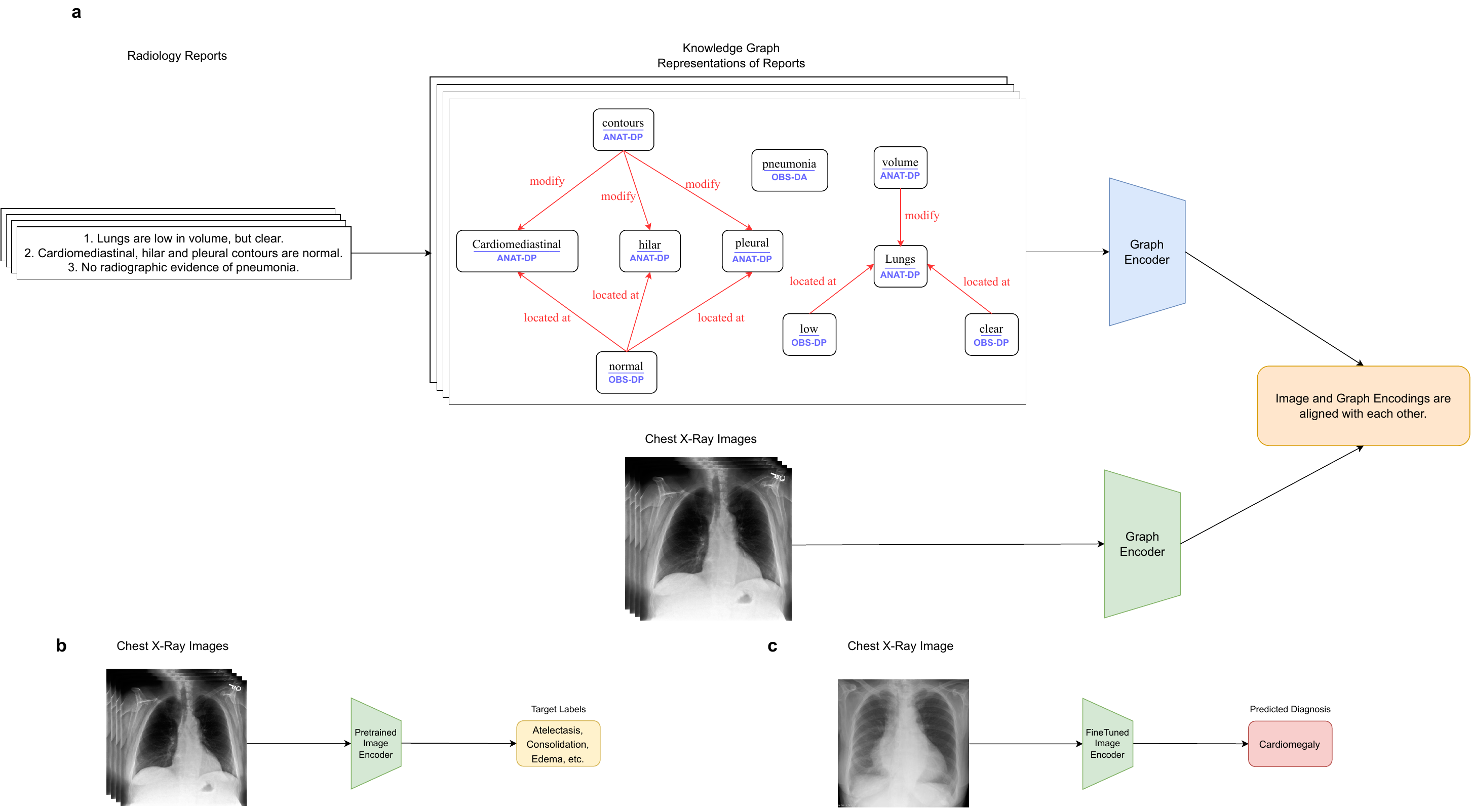}
  \caption{\textbf{a}, Chest x-ray reports are converted to more structured knowledge graph representations. The report knowledge graphs are then paired with their chest x-ray image counterparts and both sets are encoded. The image encoder and graph encoder are trained such that positive pairs are similar, while negative pairs are dissimilar. \textbf{b}, The image encoder is subsequently fine-tuned using a small set of chest x-ray and diagnosis label pairs. \textbf{c}, The fine-tuned encoder is then ready to predict diagnoses for unseen chest x-ray images.}
\end{figure*}

\section{Related Work}

\paragraph{Contrastive Learning for medical images}

Contrastive Learning for medical images is typically classified into two categories. The prevalent approach employs image-only techniques utilizing data augmentation. Positive pairs consist of altered versions of the same original image, while negative pairs involve altered versions of distinct original images. This methodology has demonstrated potential in chest X-ray diagnosis \citep{sriram2021covid, sowrirajan2021moco}. Alternatively, multiple-viewpoint methods leverage multiple scans or samples from the same patient, employing these multiple views directly as positive pairs. Another avenue involves exploiting limited annotation to group images with the same label in the embedding space, resulting in enhanced contrastive learning efficacy \citep{sellergren2022simplified}.


Image-text contrastive learning capitalizes on pairs of medical images and corresponding natural language medical reports to guide medical image representation learning. These approaches tend to yield superior medical image representations compared to image-only methodologies \citep{endo2021retrieval, huang2021gloria}.

\paragraph{Contrastive Learning using graphs and images} 

Recent research \citep{pan2022contrastive} has explored the incorporation of graphs and images for contrastive-style learning. However, the application and target task of this approach differ fundamentally from IGCL. Pan et al.'s work focuses on triplet loss, amalgamating images and text via triplets such as (Image, Relation, Image), (Image, Relation, Text), and (Text, Relation, Text) to enhance modality alignment rather than improving image representations. Additionally, the proposed technique does not directly translate to the medical field due to crucial dissimilarities in graph structure. Medical report knowledge graphs exhibit non-complete connectivity, with most reports featuring multiple disjointed components. Transferring information across these components for high-quality encoding is a formidable task, one that conventional graph encoders struggle with.

\section{Methods}

As illustrated in Figure 1, IGCL utilizes medical report knowledge graphs created through combined entity and relation extraction using information extraction models. Images are paired with their corresponding graphs and passed through their respective encoders, trained to enhance similarity for true pairs and reduce it for false pairs. The pre-trained image encoder undergoes fine-tuning using limited labeled data, making it ready for downstream tasks.

\paragraph{Contrastive Learning}
Our aim is to train IGCL such that true image-graph pairs exhibit high cosine similarity, while false pairs display low cosine similarity. During training, we sample a batch of $M$ input pairs ($I_i$, $G_i$), where $I_i$ signifies the $i$-th image, and $G_i$ corresponds to the knowledge graph. Utilizing the image encoder and graph encoder, we generate subsequent encodings denoted as ($Z^{I_i}$, $Z^{G_i}$). Our training objective encompasses two loss functions: Equation 1 illustrates the image-to-graph contrastive loss for the $i$-th pair, and Equation 2 presents the graph-to-image contrastive loss for the $i$-th pair. These two loss functions are combined through simple averaging, resulting in the loss formula for the training batch as shown in Equation 3.

\begin{equation}
    \left( \mathcal{L}_{\text{image-graph}} \right)_i = -\log \left( \frac{ \exp \left( \frac{Z^{I_i} \cdot Z^{G_i}}{ |Z^{I_i}| |Z^{G_i}|} \right) }{ \sum_{k=1}^M \frac{Z^{I_i} \cdot Z^{G_k}}{ |Z^{I_i}| |Z^{G_k}|}} \right)
\end{equation}

\begin{equation}
    \left( \mathcal{L}_{\text{graph-image}} \right)_i = -\log \left( \frac{ \exp \left( \frac{Z^{I_i} \cdot Z^{G_i}}{ |Z^{I_i}| |Z^{G_i}|} \right) }{ \sum_{k=1}^M \frac{Z^{I_k} \cdot Z^{G_i}}{ |Z^{I_k}| |Z^{G_i}|}} \right)
\end{equation}

\begin{equation}
    \mathcal{L} = \frac{1}{2M}  \sum_{i=1}^M \left( \mathcal{L}_{\text{image-graph}} \right)_i +  \left( \mathcal{L}_{\text{graph-image}} \right)_i
\end{equation}

\paragraph{Graph Encoder}
For the knowledge graph representation of the $i$-th chest x-ray report $G_i$, we define $G_i$ as ($V_i$, $E_i$), with $V_i$ and $E_i$ representing the sets of nodes and edges for $G_i$, respectively. Modern Graph Neural Networks (GNNs) adopt a neighborhood aggregation approach, where node representation is iteratively updated. The update rule is generally expressed by Equation 4.

For obtaining a graph encoding, we deploy a readout operation that combines the feature vectors of all nodes in the graph $G$, as outlined in Equation 5. Our graph neural network employs the Relational Graph Convolution Network (RGCN) for combining/aggregation layers [Schlichtkrull et al., 2018].

For Readout, we conducted experiments involving various pooling techniques, including mean pooling, min pooling, max pooling, global counterparts, and Global Attention pooling. Max pooling demonstrated superior performance in generating graph-level encodings, as corroborated by downstream task performance. A fully connected projection head is applied to the resulting encoding before it's used to train our image encoder via contrastive learning.

\paragraph{Aggregating Information Between Disconnected Components} 
The RGCN module in the graph encoder furnishes vector representations for each node in the medical report graph. This outcome enables robust encodings for each connected component within the medical report graph. However, it doesn't facilitate the flow of information across distinct disconnected components that coexist within a single medical report. The embeddings within one component remain isolated from those in disconnected components. This limitation can pose challenges, as crucial insights for comprehending one connected component might be ensconced within another graph component.

To address the aggregation of information across nodes in a disjointed graph, we propose leveraging an attention mechanism through a transformer encoder. This transformer encoder exhibits the ability to focus on different segments of input data, thereby fostering interactions among the assorted disconnected components of the graph through multi-headed attention. This empowers our model to attain a clearer understanding of the dynamics within the knowledge graph report.

Multi-head attention orchestrates the transformation of a matrix containing vector representations of nodes, as delineated in equations 7-9 below. This multi-head attention empowers the graph encoder to regulate the amalgamation of information across segments of the medical knowledge graphs, ultimately culminating in the creation of more intricate representations. This, in turn, contributes to enhanced performance on downstream tasks.

\begin{equation}
    H^{(l+1)} = concat [ \text{head}_1, \text{head}_2,\text{head}_3, ...]
\end{equation}

\begin{equation}
    \text{head}_i = \text{Attention}(H^{(l)} W_i^Q, H^{(l)} W_i^K, H^{(l)} W_i^V)
\end{equation}

\begin{equation}
    \text{Attention}(Q, K, V) = \text{softmax} \left( \frac{KQ^T}{\sqrt{\text{d}_\text{model}}} \right)
\end{equation}

Here, $Q$, $K$, $V$ are queries, keys, values, while $\text{d}_\text{model}$ is a dimension of key.

\paragraph{Image Encoder}

Our image encoder utilizes a vision transformer. While earlier medical computer vision models primarily employed deep convolutional neural networks (CNNs) like ResNet-50 [He et al., 2016], recent work suggests that pretrained transformers may yield more transferable learned representations for medical image encodings \citep{zhou2022generalized, endo2021retrieval}. Across all image-graph contrastive learning experiments, we initiate with the pretrained CLIP vision encoder. 


\section{Results}

\paragraph{IGCL can leverage information about the relationship between observations to improve performance.}

\begin{figure*}[!t]
  \label{fig:fig2}
  \centering
  \includegraphics[width=\linewidth]{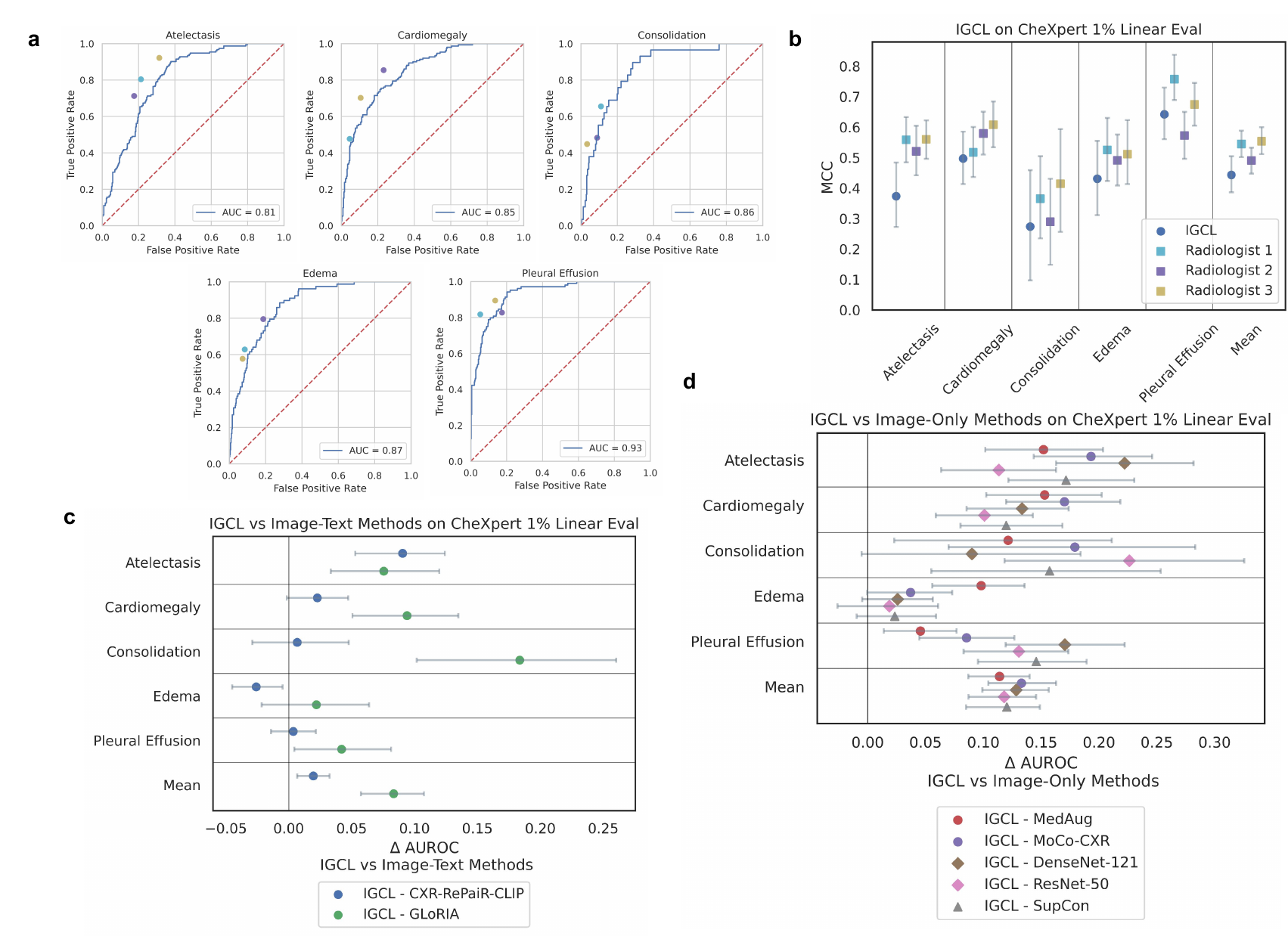}
  \caption{\textbf{a}, Receiver operating characteristic (ROC) curves for downstream evaluation of IGCL on the CheXpert dataset. The three points on each curve indicate the performance of three benchmark radiologists. \textbf{b}, Matthew's correlation coefficient (MCC) for IGCL compared with three benchmark radiologists. Error bars represent 95\% confidence intervals. \textbf{c}, Comparison of IGCL with benchmark models (CXR-RePaiR-CLIP and GLoRIA) trained via image-text contrastive learning, where models are evaluated by training a linear probe on 1\% of the CheXpert dataset. \textbf{d}, Comparison of IGCL with models pretrained via image-image contrastive learning (MedAug and MoCo-CXR), traditional image supervised learning (DenseNet and ResNet), and supervised contrastive learning (SupCon).
}
\end{figure*}

We utilize contrastive learning to enhance chest x-ray pathology classification with minimal labeled data. Our Image-Graph Contrastive Learning (IGCL) model leverages paired chest x-ray images and knowledge graph representations from the RadGraph dataset. It predicts the correspondence between x-rays and radiology report graphs, enhancing medical image representation quality through relational structure. We evaluate IGCL's medical image representations using a linear probe on 1\% of the CheXpert pathology classification dataset.

IGCL exhibits performance on par with three benchmark radiologists \citep{rajpurkar2020chexpedition} across four out of the five selected CheXpert pathologies. Comparing the receiver operating characteristic (ROC) curve of IGCL against the radiologist's performance in relation to the ground truth, we observe that IGCL surpasses radiologists when its ROC curve lies above the radiologist's operating points. As depicted in Figure 2a, IGCL closely matches radiologist performance for all pathologies except Atelectasis. Additionally, the Matthew's Correlation Coefficient (MCC) shows no significant difference between IGCL and any of the three benchmark radiologists, as demonstrated by the overlapping error bars in Figure 2b. These findings collectively affirm IGCL's capacity to achieve pathology classification proficiency comparable to that of radiologists.

Furthermore, IGCL outperforms the state-of-the-art medical image-text contrastive learning approach, CXR-RePaiR-CLIP \citep{endo2021retrieval}. IGCL matches or exceeds CXR-RePaiR-CLIP's performance on four of the five chosen pathologies, with the exception of Edema. Similarly, IGCL surpasses GLoRIA \citep{huang2021gloria}, another relevant image-text contrastive learning method, with statistical significance across all tasks except Edema. Notably, IGCL attains an AUROC of at least 0.8 for all pathologies, unlike CXR-RePaiR-CLIP, which falls below the 0.8 threshold for Atelectasis.

Moreover, the graph modality significantly contributes to IGCL's performance. In a 1\% linear evaluation context, IGCL outperforms all image-only contrastive learning methods (MedAug \citep{vu2021medaug} and MoCo-CXR \citep{sowrirajan2021moco}) and supervised learning methods (ResNet-50 \citep{he2016deep} and DenseNet-12 \citep{tan2018survey}) by over 0.1 AUROC. Together, these results underscore the pivotal role of the knowledge graph modality in augmenting downstream model efficacy.

\paragraph{Attention is the optimal way to handle disconnected graphs}

\begin{figure}[!t]
  \label{fig:fig5}
  \centering
  \includegraphics[width=\linewidth]{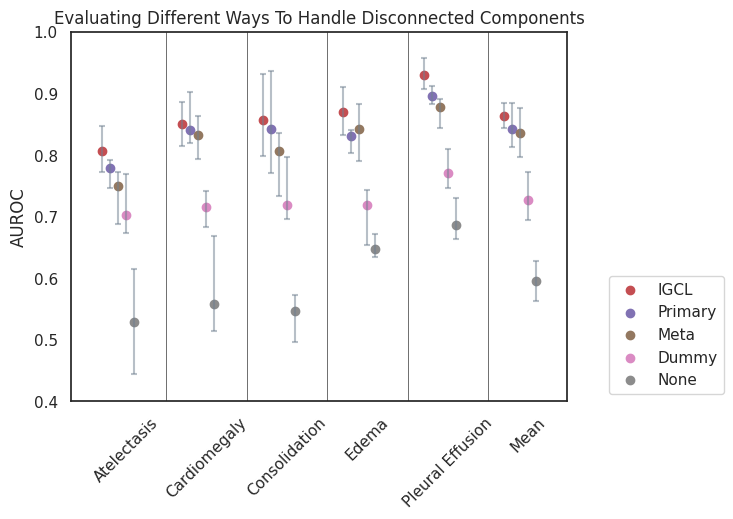}
  \caption{Comparison of Approaches to handle Disconnected Components.
}
\end{figure}

We propose adopting attention as a means to manage disconnected graph components by facilitating information exchange. An alternative approach to address this challenge would be to augment the graphs to establish connections between disconnected components. This would enable the use of conventional graph encoders. Various augmentation strategies can be employed. Initially, nodes can be linked to a single placeholder node (referred to as "dummy augmentation"). Alternatively, each node can connect to component-specific meta nodes, either forming dense interconnections (referred to as "meta augmentation") or linking to a primary node (referred to as "primary augmentation"). These approaches balance simplicity while preserving disconnected component structure. Added edges only connect disjoint components without conveying relation information, as indicated by their relation encodings being the zero vector.

We compare our proposed graph encoder with a standard RGCN encoder, incorporating each of these augmentation strategies, and assess performance without any augmentations to determine their necessity.

Figure 3 presents the results. Attention yields the best outcomes, significantly improving performance compared to graph augmentations. Meta and primary augmentations perform similarly, with dummy augmentation showing the lowest performance. This suggests that augmentations preserving relationships between connected components are more effective for graph-level encoding tasks. The control scenario, without augmentation, demonstrates the weakest performance, highlighting the distinct nature of medical report knowledge graphs compared to traditional knowledge graphs.

\begin{figure*}[!t]
  \label{fig:fig3}
  \centering
  \includegraphics[width=\linewidth]{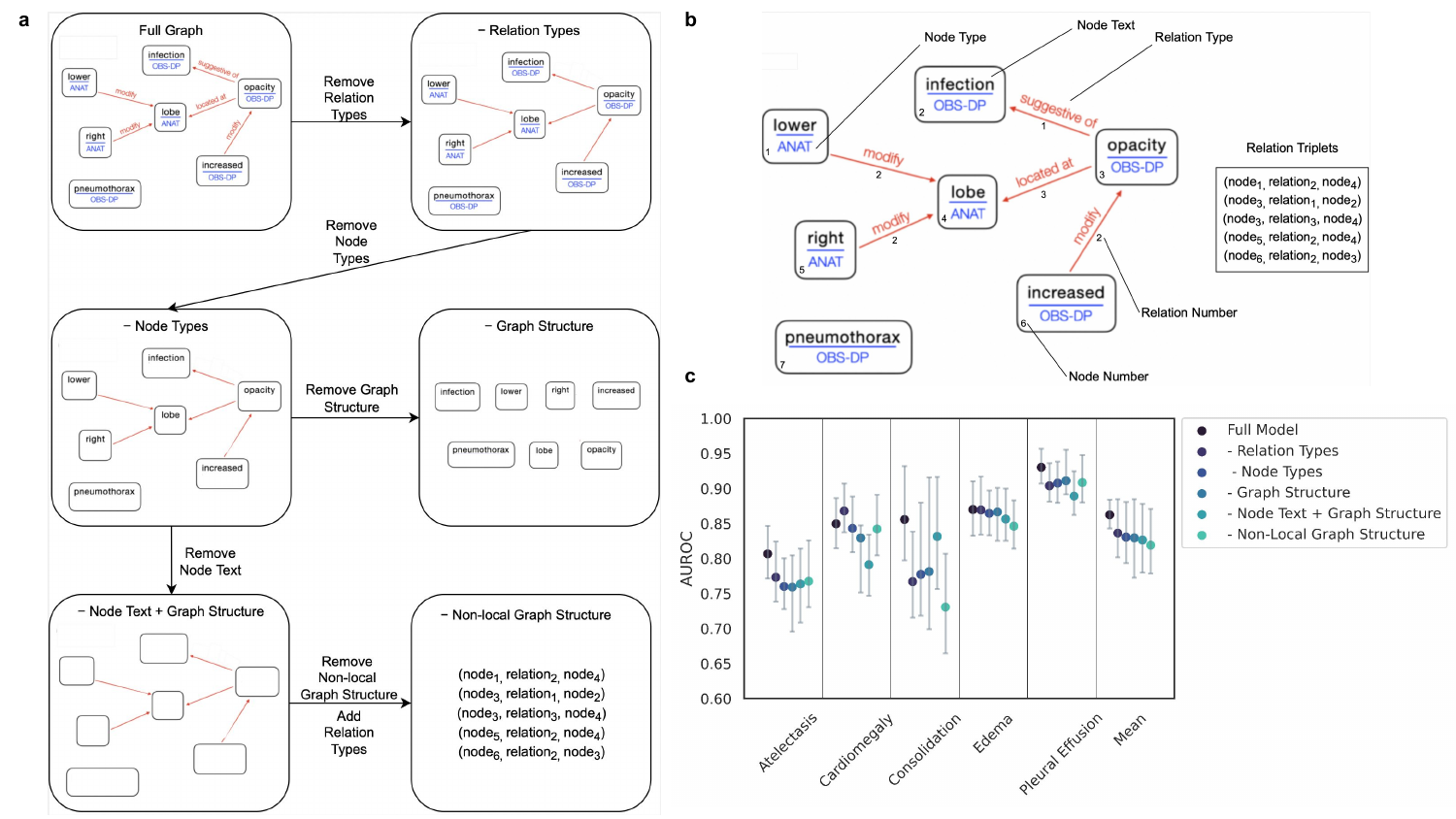}
  \caption{\textbf{a}, Ablation experimental setup. Each ablation removes information from the knowledge graph. This subfigure visually indicates how each removal relates to previous ablations. \textbf{b}, Key components of a radiology report knowledge graph evaluated via ablation. \textbf{c}, Results of the graph ablation experiments.
}
\end{figure*}

\paragraph{Simplifying Graph Structure or Removing Graphs impairs downstream performance.}

We delve into the significance of different knowledge graph components for learning high-quality medical image representations. Through ablation experiments, we eliminate key elements from the knowledge graph (illustrated in Figures 3a and 3b) to assess the contributions of node text, node type, relation information, and graph structure. We also evaluate the importance of both local and global graph structure, examining whether adequate representations can be achieved with only relation triplets.

Model selections for each ablation are based on their ability to incorporate the given information. For instance, RotatE \citep{sun2019rotate} is chosen for the non-local graph structure ablation due to its effective utilization of relation triplets. With the exception of the no graph structure and no non-local graph structure models, we utilize a GCN graph encoder \citep{zhang2019graph} as the base, as the ablated models do not differentiate between various relation types. The no-graph structure model encodes graphs as the mean BERT embeddings of node text \citep{devlin2018bert}. The final model encodes graphs using the mean RotatE embeddings of relation triplets.

Figure 3c illustrates the results, showing that removing any piece of information from the knowledge graph leads to a significant reduction in downstream task performance, causing the mean AUC to drop from 0.864 to 0.819. The most pronounced decline occurs when relation types are removed from the graph. This information loss substantially outweighs the impact of the chosen graph encoder. For instance, replacing the RGCN \citep{schlichtkrull2018modeling} network with a GAT \citep{velivckovic2017graph} network in our architecture yields similar performance, with RGCN achieving a mean AUC of 0.863, and GAT achieving a mean AUC of 0.862.

\begin{figure*}[!t]
  \label{fig:fig4}
  \centering
  \includegraphics[width=\linewidth]{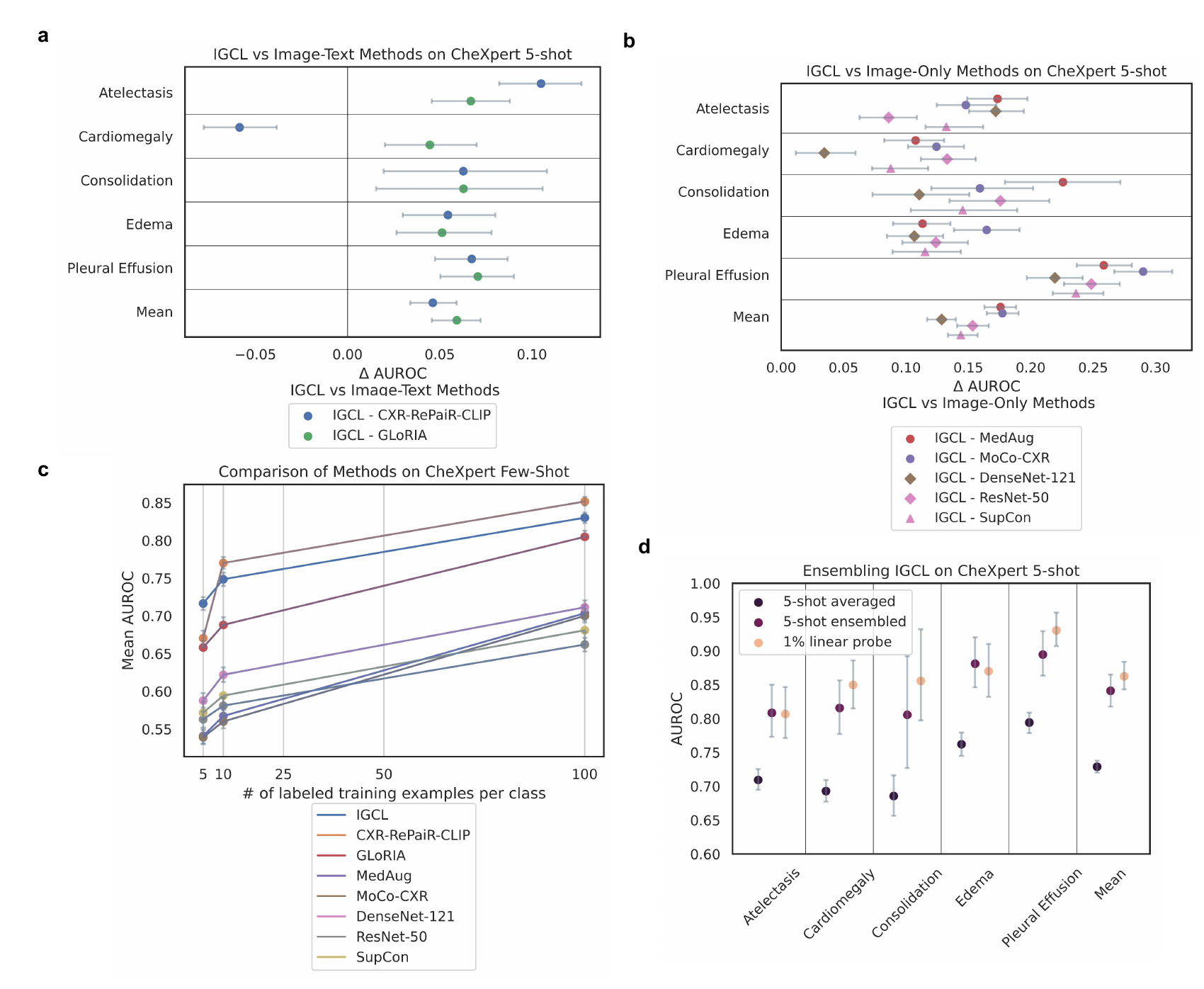}
  \caption{\textbf{a}, Comparison of IGCL with models trained via image-text contrastive learning. Models are evaluated by training a 5-shot linear probe on the CheXpert dataset. \textbf{b}, Comparison of IGCL with models pre-trained via image only contrastive learning (MedAug and MoCo-CXR) and traditional image supervised learning (DenseNet and ResNet). \textbf{c}, Comparison of IGCL performance across 5, 10, 20, and 50-shot CheXpert evaluation settings. d, Ensembles of 10 independently trained 5-shot IGCL models vs 1\% linear probe.
}
\end{figure*}

\paragraph{Image-graph contrastive learning demonstrates superior label efficiency.}

Label efficiency holds paramount importance in the medical field due to the limited size of medical datasets and the challenges of obtaining clinician annotations. To evaluate label efficiency, we subject IGCL and baseline models to few-shot settings: 5-shot, 10-shot, 20-shot, and 50-shot. In these scenarios, a linear probe is trained with only 5, 10, 20, or 50 images per label. For comparison, in the 1\% linear evaluation, the model trains on what is equivalent to 372-shot if all labels had equal likelihood.

In the 5-shot setting, IGCL significantly outperforms all image-text and image-only baseline methods in terms of mean AUROC. Figures 4a and 4b provide an overview of these outcomes. Notably, IGCL surpasses CXR-RePaiR-CLIP \citep{endo2021retrieval}, GLoRIA \citep{huang2021gloria}, and all image-only \citep{he2016deep, huang2017densely, vu2021medaug, sowrirajan2021moco} methods by even wider margins compared to the 1\% linear evaluation. This results in mean AUROC values 6\% higher than image-text methods and 18\% higher than image-only methods. Thus, IGCL not only benefits from exploiting the relationship between observations for enhanced performance, but it also effectively harnesses labeled data in few-shot settings. This is further emphasized by Figure 4c, demonstrating that 5-shot IGCL significantly outperforms image-only baseline methods trained with 50-shot, nearly equating the performance of image-only methods under linear evaluation. Consequently, IGCL demonstrates over 10 times greater label efficiency than image-only methods in few-shot settings.

These few-shot experiments also unveil the effectiveness of synthesizing multiple models in limited-data scenarios. Figures 4a, b, and c showcase averages across 10 independently trained models for each architecture. Ensembling these 10 models by deriving the median consensus of predicted probabilities leads to a statistically significant improvement in IGCL model performance. Impressively, mean AUROC improves by over 0.1 through ensembling, aligning the ensemble's performance closely with the 1\% linear probe, as illustrated by the overlapping error bars in Figure 4d. This underscores the efficacy of ensembling multiple few-shot IGCL models, yielding markedly improved disease diagnostic capabilities.

\section{Discussion}

This study highlights the effectiveness of knowledge graphs in enhancing medical image embeddings. Our approach is assessed through improved performance in five downstream tasks. Furthermore, our research explores the broader potential of knowledge graph modality, illuminating optimal machine learning pipeline design for graph structure utilization. Specifically, our innovative attention-based method addresses the challenge of disconnected components in radiology report knowledge graphs, a feature uncommon in traditional graph structures. This work contributes in multiple ways: introducing a superior image-graph contrastive pretraining strategy and proposing an efficient graph encoder for handling disconnected knowledge graphs. Our work can be expanded in the future beyond disease detection to enable the generation of free-text radiology reports or apply to other imaging modalities.

\paragraph{IGCL outperforms image-text contrastive learning methods by substantial margins.}

A significant revelation is the superiority of image-graph contrastive learning over state-of-the-art image-text methods, both in 1\% linear evaluation and few-shot scenarios. IGCL consistently achieves significantly higher mean AUROC values than the best-performing baseline, CXR-RePaiR-CLIP, in both 1\% and 5-shot experiments. Especially in the 5-shot context, IGCL surpasses all other baseline methods significantly, except for CXR-RePaiR-CLIP, which performs similarly. This suggests that the graph modality's guidance in learning image embeddings requires fewer labeled examples for competitive performance against image-text approaches, but this advantage diminishes as the number of available labels increases.

The fact that IGCL outperforms CXR-RePaiR-CLIP is significant, given their substantial similarity. Both models use contrastive pretraining on MIMIC-CXR images with the same loss function and employ identical vision transformer architectures for image encoding. The key difference lies in the data used for pretraining: CXR-RePaiR-CLIP uses raw radiology reports, while IGCL uses auto-generated RadGraph knowledge graphs from these reports. This direct comparison highlights the advantage of using knowledge graphs over textual reports during contrastive pretraining, leading to superior image embeddings. This finding carries notable implications, especially in the context of widely used image-text contrastive learning models like CLIP \citep{radford2021learning}. Image-graph contrastive learning introduces potential for positive enhancements in image-text pretrained applications.

\paragraph{Radiology report knowledge graph learning differs from traditional graph learning.}
A significant contribution lies in our exploration of creating graph-level encodings for radiology report knowledge graphs, which significantly differ in structure and task nature compared to traditional graphs in fields like chemistry and biology. Existing datasets in such fields, including AIFB \citep{bloehdorn2007kernel}, MUTAG45 \citep{debnath1991structure}, BGS \citep{ristoski2016collection}, AM \citep{de2012supporting}, and MD17 \citep{chmiela2017machine}, mainly involve large-scale graphs targeting entity classification or link prediction tasks. Graph-level datasets such as ALCHEMY \citep{chen2019alchemy}, ZINC \citep{chen2019alchemy}, and QM9 \citep{ruddigkeit2012enumeration, ramakrishnan2014quantum} center on molecular tasks, aiming to predict key molecular properties from graph representations. These tasks, focusing on the molecule level, often involve graphs with single connected components and no edge information. Notably, IGCL stands out by working with numerous small, heterogeneous graphs for graph classification, comprising multiple disconnected components.

\paragraph{Each component in the medical report knowledge graph is crucial.}

Our ablation studies offer valuable insights into each knowledge graph component's contributions to learning high-quality image representations. Removing any component from the knowledge graph significantly reduces performance in downstream tasks. The most substantial decline occurs when relation information is omitted, leading to the loss of contextual information for connecting nodes. Similarly, removing node text information causes a noticeable performance drop, indicating that having access to report information enhances performance, regardless of textual information ordering within radiology reports. The impact of removing relational information is also evident; on average, its absence leads to a reduction of performance.

\newpage

\bibliography{khanna23}

\begin{thebibliography}{39}
\providecommand{\natexlab}[1]{#1}
\providecommand{\url}[1]{\texttt{#1}}
\expandafter\ifx\csname urlstyle\endcsname\relax
  \providecommand{\doi}[1]{doi: #1}\else
  \providecommand{\doi}{doi: \begingroup \urlstyle{rm}\Url}\fi

\bibitem[Abràmoff et~al.(2016)Abràmoff, Lou, Erginay, Clarida, Amelon, Folk,
  and Niemeijer]{10.1167/iovs.16-19964}
Michael~David Abràmoff, Yiyue Lou, Ali Erginay, Warren Clarida, Ryan Amelon,
  James~C. Folk, and Meindert Niemeijer.
\newblock {Improved Automated Detection of Diabetic Retinopathy on a Publicly
  Available Dataset Through Integration of Deep Learning}.
\newblock \emph{Investigative Ophthalmology \& Visual Science}, 57\penalty0
  (13):\penalty0 5200--5206, 10 2016.
\newblock ISSN 1552-5783.
\newblock \doi{10.1167/iovs.16-19964}.
\newblock URL \url{https://doi.org/10.1167/iovs.16-19964}.

\bibitem[Bloehdorn and Sure(2007)]{bloehdorn2007kernel}
Stephan Bloehdorn and York Sure.
\newblock Kernel methods for mining instance data in ontologies.
\newblock In \emph{The Semantic Web: 6th International Semantic Web Conference,
  2nd Asian Semantic Web Conference, ISWC 2007+ ASWC 2007, Busan, Korea,
  November 11-15, 2007. Proceedings}, pages 58--71. Springer, 2007.

\bibitem[Chen et~al.(2019{\natexlab{a}})Chen, Chen, Hsieh, Lee, Liao, Liao,
  Liu, Qiu, Sun, Tang, et~al.]{chen2019alchemy}
Guangyong Chen, Pengfei Chen, Chang-Yu Hsieh, Chee-Kong Lee, Benben Liao,
  Renjie Liao, Weiwen Liu, Jiezhong Qiu, Qiming Sun, Jie Tang, et~al.
\newblock Alchemy: A quantum chemistry dataset for benchmarking ai models.
\newblock \emph{arXiv preprint arXiv:1906.09427}, 2019{\natexlab{a}}.

\bibitem[Chen et~al.(2019{\natexlab{b}})Chen, Bentley, Mori, Misawa, Fujiwara,
  and Rueckert]{chen2019self}
Liang Chen, Paul Bentley, Kensaku Mori, Kazunari Misawa, Michitaka Fujiwara,
  and Daniel Rueckert.
\newblock Self-supervised learning for medical image analysis using image
  context restoration.
\newblock \emph{Medical image analysis}, 58:\penalty0 101539,
  2019{\natexlab{b}}.

\bibitem[Chen et~al.(2020)Chen, Kornblith, Norouzi, and Hinton]{chen2020simple}
Ting Chen, Simon Kornblith, Mohammad Norouzi, and Geoffrey Hinton.
\newblock A simple framework for contrastive learning of visual
  representations.
\newblock In \emph{International conference on machine learning}, pages
  1597--1607. PMLR, 2020.

\bibitem[Chmiela et~al.(2017)Chmiela, Tkatchenko, Sauceda, Poltavsky,
  Sch{\"u}tt, and M{\"u}ller]{chmiela2017machine}
Stefan Chmiela, Alexandre Tkatchenko, Huziel~E Sauceda, Igor Poltavsky,
  Kristof~T Sch{\"u}tt, and Klaus-Robert M{\"u}ller.
\newblock Machine learning of accurate energy-conserving molecular force
  fields.
\newblock \emph{Science advances}, 3\penalty0 (5):\penalty0 e1603015, 2017.

\bibitem[De~Boer et~al.(2012)De~Boer, Wielemaker, Van~Gent, Hildebrand, Isaac,
  Van~Ossenbruggen, and Schreiber]{de2012supporting}
Victor De~Boer, Jan Wielemaker, Judith Van~Gent, Michiel Hildebrand, Antoine
  Isaac, Jacco Van~Ossenbruggen, and Guus Schreiber.
\newblock Supporting linked data production for cultural heritage institutes:
  the amsterdam museum case study.
\newblock In \emph{The Semantic Web: Research and Applications: 9th Extended
  Semantic Web Conference, ESWC 2012, Heraklion, Crete, Greece, May 27-31,
  2012. Proceedings 9}, pages 733--747. Springer, 2012.

\bibitem[De~Fauw et~al.(2018)De~Fauw, Ledsam, Romera-Paredes, Nikolov, Tomasev,
  Blackwell, Askham, Glorot, O’Donoghue, Visentin, et~al.]{de2018clinically}
Jeffrey De~Fauw, Joseph~R Ledsam, Bernardino Romera-Paredes, Stanislav Nikolov,
  Nenad Tomasev, Sam Blackwell, Harry Askham, Xavier Glorot, Brendan
  O’Donoghue, Daniel Visentin, et~al.
\newblock Clinically applicable deep learning for diagnosis and referral in
  retinal disease.
\newblock \emph{Nature medicine}, 24\penalty0 (9):\penalty0 1342--1350, 2018.

\bibitem[Debnath et~al.(1991)Debnath, Lopez~de Compadre, Debnath, Shusterman,
  and Hansch]{debnath1991structure}
Asim~Kumar Debnath, Rosa~L Lopez~de Compadre, Gargi Debnath, Alan~J Shusterman,
  and Corwin Hansch.
\newblock Structure-activity relationship of mutagenic aromatic and
  heteroaromatic nitro compounds. correlation with molecular orbital energies
  and hydrophobicity.
\newblock \emph{Journal of medicinal chemistry}, 34\penalty0 (2):\penalty0
  786--797, 1991.

\bibitem[Devlin et~al.(2018)Devlin, Chang, Lee, and Toutanova]{devlin2018bert}
Jacob Devlin, Ming-Wei Chang, Kenton Lee, and Kristina Toutanova.
\newblock Bert: Pre-training of deep bidirectional transformers for language
  understanding.
\newblock \emph{arXiv preprint arXiv:1810.04805}, 2018.

\bibitem[Endo et~al.(2021)Endo, Krishnan, Krishna, Ng, and
  Rajpurkar]{endo2021retrieval}
Mark Endo, Rayan Krishnan, Viswesh Krishna, Andrew~Y Ng, and Pranav Rajpurkar.
\newblock Retrieval-based chest x-ray report generation using a pre-trained
  contrastive language-image model.
\newblock In \emph{Machine Learning for Health}, pages 209--219. PMLR, 2021.

\bibitem[He et~al.(2016)He, Zhang, Ren, and Sun]{he2016deep}
Kaiming He, Xiangyu Zhang, Shaoqing Ren, and Jian Sun.
\newblock Deep residual learning for image recognition.
\newblock In \emph{Proceedings of the IEEE conference on computer vision and
  pattern recognition}, pages 770--778, 2016.

\bibitem[Huang et~al.(2017)Huang, Liu, Van Der~Maaten, and
  Weinberger]{huang2017densely}
Gao Huang, Zhuang Liu, Laurens Van Der~Maaten, and Kilian~Q Weinberger.
\newblock Densely connected convolutional networks.
\newblock In \emph{Proceedings of the IEEE conference on computer vision and
  pattern recognition}, pages 4700--4708, 2017.

\bibitem[Huang et~al.(2021)Huang, Shen, Lungren, and Yeung]{huang2021gloria}
Shih-Cheng Huang, Liyue Shen, Matthew~P Lungren, and Serena Yeung.
\newblock Gloria: A multimodal global-local representation learning framework
  for label-efficient medical image recognition.
\newblock In \emph{Proceedings of the IEEE/CVF International Conference on
  Computer Vision}, pages 3942--3951, 2021.

\bibitem[Irvin et~al.(2019)Irvin, Rajpurkar, Ko, Yu, Ciurea-Ilcus, Chute,
  Marklund, Haghgoo, Ball, Shpanskaya, et~al.]{irvin2019chexpert}
Jeremy Irvin, Pranav Rajpurkar, Michael Ko, Yifan Yu, Silviana Ciurea-Ilcus,
  Chris Chute, Henrik Marklund, Behzad Haghgoo, Robyn Ball, Katie Shpanskaya,
  et~al.
\newblock Chexpert: A large chest radiograph dataset with uncertainty labels
  and expert comparison.
\newblock In \emph{Proceedings of the AAAI conference on artificial
  intelligence}, volume~33, pages 590--597, 2019.

\bibitem[Jain et~al.(2021)Jain, Agrawal, Saporta, Truong, Duong, Bui, Chambon,
  Zhang, Lungren, Ng, Langlotz, Rajpurkar, and Rajpurkar]{radgraph-jain}
Saahil Jain, Ashwin Agrawal, Adriel Saporta, Steven Truong, Du~Nguyen Duong,
  Tan Bui, Pierre Chambon, Yuhao Zhang, Matthew Lungren, Andrew Ng, Curtis
  Langlotz, Pranav Rajpurkar, and Pranav Rajpurkar.
\newblock Radgraph: Extracting clinical entities and relations from radiology
  reports.
\newblock In J.~Vanschoren and S.~Yeung, editors, \emph{Proceedings of the
  Neural Information Processing Systems Track on Datasets and Benchmarks},
  volume~1, 2021.

\bibitem[Ke et~al.(2021)Ke, Ellsworth, Banerjee, Ng, and
  Rajpurkar]{ke2021chextransfer}
Alexander Ke, William Ellsworth, Oishi Banerjee, Andrew~Y Ng, and Pranav
  Rajpurkar.
\newblock Chextransfer: performance and parameter efficiency of imagenet models
  for chest x-ray interpretation.
\newblock In \emph{Proceedings of the conference on health, inference, and
  learning}, pages 116--124, 2021.

\bibitem[Misra and Maaten(2020)]{misra2020self}
Ishan Misra and Laurens van~der Maaten.
\newblock Self-supervised learning of pretext-invariant representations.
\newblock In \emph{Proceedings of the IEEE/CVF conference on computer vision
  and pattern recognition}, pages 6707--6717, 2020.

\bibitem[Pan et~al.(2022)Pan, Ye, Han, Song, and Huang]{pan2022contrastive}
Xuran Pan, Tianzhu Ye, Dongchen Han, Shiji Song, and Gao Huang.
\newblock Contrastive language-image pre-training with knowledge graphs.
\newblock \emph{arXiv preprint arXiv:2210.08901}, 2022.

\bibitem[Radford et~al.(2021)Radford, Kim, Hallacy, Ramesh, Goh, Agarwal,
  Sastry, Askell, Mishkin, Clark, et~al.]{radford2021learning}
Alec Radford, Jong~Wook Kim, Chris Hallacy, Aditya Ramesh, Gabriel Goh,
  Sandhini Agarwal, Girish Sastry, Amanda Askell, Pamela Mishkin, Jack Clark,
  et~al.
\newblock Learning transferable visual models from natural language
  supervision.
\newblock In \emph{International conference on machine learning}, pages
  8748--8763. PMLR, 2021.

\bibitem[Rajpurkar et~al.(2018)Rajpurkar, Irvin, Ball, Zhu, Yang, Mehta, Duan,
  Ding, Bagul, Langlotz, et~al.]{rajpurkar2018deep}
Pranav Rajpurkar, Jeremy Irvin, Robyn~L Ball, Kaylie Zhu, Brandon Yang, Hershel
  Mehta, Tony Duan, Daisy Ding, Aarti Bagul, Curtis~P Langlotz, et~al.
\newblock Deep learning for chest radiograph diagnosis: A retrospective
  comparison of the chexnext algorithm to practicing radiologists.
\newblock \emph{PLoS medicine}, 15\penalty0 (11):\penalty0 e1002686, 2018.

\bibitem[Rajpurkar et~al.(2020{\natexlab{a}})Rajpurkar, Joshi, Pareek, Chen,
  Kiani, Irvin, Ng, and Lungren]{rajpurkar2020chexpedition}
Pranav Rajpurkar, Anirudh Joshi, Anuj Pareek, Phil Chen, Amirhossein Kiani,
  Jeremy Irvin, Andrew~Y Ng, and Matthew~P Lungren.
\newblock Chexpedition: investigating generalization challenges for translation
  of chest x-ray algorithms to the clinical setting.
\newblock \emph{arXiv preprint arXiv:2002.11379}, 2020{\natexlab{a}}.

\bibitem[Rajpurkar et~al.(2020{\natexlab{b}})Rajpurkar, O’Connell, Schechter,
  Asnani, Li, Kiani, Ball, Mendelson, Maartens, van Hoving,
  et~al.]{rajpurkar2020chexaid}
Pranav Rajpurkar, Chloe O’Connell, Amit Schechter, Nishit Asnani, Jason Li,
  Amirhossein Kiani, Robyn~L Ball, Marc Mendelson, Gary Maartens, Dani{\"e}l~J
  van Hoving, et~al.
\newblock Chexaid: deep learning assistance for physician diagnosis of
  tuberculosis using chest x-rays in patients with hiv.
\newblock \emph{NPJ digital medicine}, 3\penalty0 (1):\penalty0 115,
  2020{\natexlab{b}}.

\bibitem[Ramakrishnan et~al.(2014)Ramakrishnan, Dral, Rupp, and
  Von~Lilienfeld]{ramakrishnan2014quantum}
Raghunathan Ramakrishnan, Pavlo~O Dral, Matthias Rupp, and O~Anatole
  Von~Lilienfeld.
\newblock Quantum chemistry structures and properties of 134 kilo molecules.
\newblock \emph{Scientific data}, 1\penalty0 (1):\penalty0 1--7, 2014.

\bibitem[Ristoski et~al.(2016)Ristoski, De~Vries, and
  Paulheim]{ristoski2016collection}
Petar Ristoski, Gerben Klaas~Dirk De~Vries, and Heiko Paulheim.
\newblock A collection of benchmark datasets for systematic evaluations of
  machine learning on the semantic web.
\newblock In \emph{The Semantic Web--ISWC 2016: 15th International Semantic Web
  Conference, Kobe, Japan, October 17--21, 2016, Proceedings, Part II 15},
  pages 186--194. Springer, 2016.

\bibitem[Ruddigkeit et~al.(2012)Ruddigkeit, Van~Deursen, Blum, and
  Reymond]{ruddigkeit2012enumeration}
Lars Ruddigkeit, Ruud Van~Deursen, Lorenz~C Blum, and Jean-Louis Reymond.
\newblock Enumeration of 166 billion organic small molecules in the chemical
  universe database gdb-17.
\newblock \emph{Journal of chemical information and modeling}, 52\penalty0
  (11):\penalty0 2864--2875, 2012.

\bibitem[Schlichtkrull et~al.(2018)Schlichtkrull, Kipf, Bloem, Van Den~Berg,
  Titov, and Welling]{schlichtkrull2018modeling}
Michael Schlichtkrull, Thomas~N Kipf, Peter Bloem, Rianne Van Den~Berg, Ivan
  Titov, and Max Welling.
\newblock Modeling relational data with graph convolutional networks.
\newblock In \emph{The Semantic Web: 15th International Conference, ESWC 2018,
  Heraklion, Crete, Greece, June 3--7, 2018, Proceedings 15}, pages 593--607.
  Springer, 2018.

\bibitem[Sellergren et~al.(2022)Sellergren, Chen, Nabulsi, Li, Maschinot,
  Sarna, Huang, Lau, Kalidindi, Etemadi, et~al.]{sellergren2022simplified}
Andrew~B Sellergren, Christina Chen, Zaid Nabulsi, Yuanzhen Li, Aaron
  Maschinot, Aaron Sarna, Jenny Huang, Charles Lau, Sreenivasa~Raju Kalidindi,
  Mozziyar Etemadi, et~al.
\newblock Simplified transfer learning for chest radiography models using less
  data.
\newblock \emph{Radiology}, 305\penalty0 (2):\penalty0 454--465, 2022.

\bibitem[Shih et~al.(2019)Shih, Wu, Halabi, Kohli, Prevedello, Cook, Sharma,
  Amorosa, Arteaga, Galperin-Aizenberg, et~al.]{shih2019augmenting}
George Shih, Carol~C Wu, Safwan~S Halabi, Marc~D Kohli, Luciano~M Prevedello,
  Tessa~S Cook, Arjun Sharma, Judith~K Amorosa, Veronica Arteaga, Maya
  Galperin-Aizenberg, et~al.
\newblock Augmenting the national institutes of health chest radiograph dataset
  with expert annotations of possible pneumonia.
\newblock \emph{Radiology: Artificial Intelligence}, 1\penalty0 (1):\penalty0
  e180041, 2019.

\bibitem[Sowrirajan et~al.(2021)Sowrirajan, Yang, Ng, and
  Rajpurkar]{sowrirajan2021moco}
Hari Sowrirajan, Jingbo Yang, Andrew~Y Ng, and Pranav Rajpurkar.
\newblock Moco pretraining improves representation and transferability of chest
  x-ray models.
\newblock In \emph{Medical Imaging with Deep Learning}, pages 728--744. PMLR,
  2021.

\bibitem[Sriram et~al.(2021)Sriram, Muckley, Sinha, Shamout, Pineau, Geras,
  Azour, Aphinyanaphongs, Yakubova, and Moore]{sriram2021covid}
Anuroop Sriram, Matthew Muckley, Koustuv Sinha, Farah Shamout, Joelle Pineau,
  Krzysztof~J Geras, Lea Azour, Yindalon Aphinyanaphongs, Nafissa Yakubova, and
  William Moore.
\newblock Covid-19 prognosis via self-supervised representation learning and
  multi-image prediction.
\newblock \emph{arXiv preprint arXiv:2101.04909}, 2021.

\bibitem[Sun et~al.(2019)Sun, Deng, Nie, and Tang]{sun2019rotate}
Zhiqing Sun, Zhi-Hong Deng, Jian-Yun Nie, and Jian Tang.
\newblock Rotate: Knowledge graph embedding by relational rotation in complex
  space.
\newblock \emph{arXiv preprint arXiv:1902.10197}, 2019.

\bibitem[Tan et~al.(2018)Tan, Sun, Kong, Zhang, Yang, and Liu]{tan2018survey}
Chuanqi Tan, Fuchun Sun, Tao Kong, Wenchang Zhang, Chao Yang, and Chunfang Liu.
\newblock A survey on deep transfer learning.
\newblock In \emph{Artificial Neural Networks and Machine Learning--ICANN 2018:
  27th International Conference on Artificial Neural Networks, Rhodes, Greece,
  October 4-7, 2018, Proceedings, Part III 27}, pages 270--279. Springer, 2018.

\bibitem[Vaswani et~al.(2017)Vaswani, Shazeer, Parmar, Uszkoreit, Jones, Gomez,
  Kaiser, and Polosukhin]{vaswani2017attention}
Ashish Vaswani, Noam Shazeer, Niki Parmar, Jakob Uszkoreit, Llion Jones,
  Aidan~N Gomez, {\L}ukasz Kaiser, and Illia Polosukhin.
\newblock Attention is all you need.
\newblock \emph{Advances in neural information processing systems}, 30, 2017.

\bibitem[Veli{\v{c}}kovi{\'c} et~al.(2017)Veli{\v{c}}kovi{\'c}, Cucurull,
  Casanova, Romero, Lio, and Bengio]{velivckovic2017graph}
Petar Veli{\v{c}}kovi{\'c}, Guillem Cucurull, Arantxa Casanova, Adriana Romero,
  Pietro Lio, and Yoshua Bengio.
\newblock Graph attention networks.
\newblock \emph{arXiv preprint arXiv:1710.10903}, 2017.

\bibitem[Vu et~al.(2021)Vu, Wang, Balachandar, Liu, Ng, and
  Rajpurkar]{vu2021medaug}
Yen Nhi~Truong Vu, Richard Wang, Niranjan Balachandar, Can Liu, Andrew~Y Ng,
  and Pranav Rajpurkar.
\newblock Medaug: Contrastive learning leveraging patient metadata improves
  representations for chest x-ray interpretation.
\newblock In \emph{Machine Learning for Healthcare Conference}, pages 755--769.
  PMLR, 2021.

\bibitem[Wang et~al.(2020)Wang, Lin, and Wong]{wang2020covid}
Linda Wang, Zhong~Qiu Lin, and Alexander Wong.
\newblock Covid-net: A tailored deep convolutional neural network design for
  detection of covid-19 cases from chest x-ray images.
\newblock \emph{Scientific reports}, 10\penalty0 (1):\penalty0 1--12, 2020.

\bibitem[Zhang et~al.(2019)Zhang, Tong, Xu, and Maciejewski]{zhang2019graph}
Si~Zhang, Hanghang Tong, Jiejun Xu, and Ross Maciejewski.
\newblock Graph convolutional networks: a comprehensive review.
\newblock \emph{Computational Social Networks}, 6\penalty0 (1):\penalty0 1--23,
  2019.

\bibitem[Zhou et~al.(2022)Zhou, Chen, Zhang, Luo, Wang, and
  Yu]{zhou2022generalized}
Hong-Yu Zhou, Xiaoyu Chen, Yinghao Zhang, Ruibang Luo, Liansheng Wang, and
  Yizhou Yu.
\newblock Generalized radiograph representation learning via cross-supervision
  between images and free-text radiology reports.
\newblock \emph{Nature Machine Intelligence}, 4\penalty0 (1):\penalty0 32--40,
  2022.

\end{thebibliography}

\end{document}